\documentclass[twocolumn,prb,aps,superscriptaddress,showpacs,floatfix]{revtex4-2}
\usepackage{amssymb}
\usepackage{dcolumn}
\usepackage{epsfig}
\usepackage{graphicx}
\usepackage{epstopdf}
\usepackage{float}
\usepackage[tbtags]{amsmath}
\newlength{\figwidth}
\newlength{\figwidthb}
\begin{document}

\title{Electronic Excitations of $\alpha-$\rm Fe$_{2}$O$_{3}$ Heteroepitaxial Films Measured by Resonant Inelastic X-Ray Scattering at the Fe $L$-edge}

\author{David~S. Ellis}
\affiliation{Department of Materials Science and Engineering, Technion-Israel Institute of Technology, Haifa 32000, Israel}
\author{Ru-Pan Wang}
\affiliation{Department of Physics, University of Hamburg, Luruper Chaussee 149, 22761 Hamburg, Germany}
\author{Deniz Wong}
\affiliation{Dynamics and Transport in Quantum Materials, Helmholtz-Zentrum Berlin für Materialen und Energie, Albert-Einstein-Strasse 15, 12489 Berlin, Germany}
\author{Jason K. Cooper}
\affiliation{Chemical Sciences Division, Lawrence Berkeley National Laboratory, 1 Cyclotron Road, Berkeley, CA 94720,USA}
\author{Christian Schulz}
\affiliation{Helmholtz-Zentrum Berlin für Materialen und Energie, Albert-Einstein-Strasse 15, 12489 Berlin, Germany}
\author{Yi-De Chuang}
\affiliation{Advanced Light Source, Lawrence Berkeley National Laboratory, 1 Cyclotron Road, Berkeley, CA 94720,USA}
\author{Yifat Piekner}
\affiliation{The Nancy \& Stephen Grand Technion Energy Program (GTEP), Technion-Israel Institute of Technology, Haifa 32000, Israel}
\author{Daniel~A. Grave}
\affiliation{Department of Materials Science and Engineering, Technion-Israel Institute of Technology, Haifa 32000, Israel}
\affiliation{Department of Materials Engineering and Ilse Katz Institute for Nanoscale Science and Technology, Ben Gurion University of the Negev, Be’er Sheva 8410501, Israel}
\author{Markus Schleuning}
\affiliation{Institute for Solar Fuels, Helmholtz-Zentrum Berlin für Materialien und Energie GmbH, Hahn-Meitner-Platz 1, 14109 Berlin, Germany}
\author{Dennis Friedrich}
\affiliation{Institute for Solar Fuels, Helmholtz-Zentrum Berlin für Materialien und Energie GmbH, Hahn-Meitner-Platz 1, 14109 Berlin, Germany}
\author{Frank M. F. de Groot}
\affiliation{Department of Inorganic Chemistry and Catalysis, Debye Institute of Nanomaterials Science, Utrecht University, Universiteitsweg 99, 3584 CG Utrecht, Netherlands}
\author{Avner Rothschild}
\affiliation{Department of Materials Science and Engineering, Technion-Israel Institute of Technology, Haifa 32000, Israel}

\email{avner@mt.technion.ac.il}

\date{\today }

\begin{abstract}
Resonant Inelastic X-Ray Scattering (RIXS) spectra of hematite ($\alpha$-$\rm Fe{_2}O_{3})$ were measured at the Fe $L_{3}$-edge for heteroepitaxial thin films which were undoped and doped with 1\% Ti, Sn or Zn, in the energy loss range in excess of 1 eV to study electronic transitions. The spectra were measured for several momentum transfers (\textbf{q}), conducted at both low temperature ($T$=14K) and room temperature. While we can not rule out dispersive features possibly owing to propagating excitations, the coarse envelopes of the general spectra did not appreciably change shape with \textbf{q}, implying that the bulk of the observed $L$-edge RIXS intensity originates from (mostly) non-dispersive ligand field (LF) excitations. Summing the RIXS spectra over \textbf{q} and comparing the results at $T$=14 K to those at $T$=300 K, revealed pronounced temperature effects, including an intensity change and energy shift of the $\sim$1.4 eV peak, a broadband intensity increase of the 3-4 eV range, and higher energy features. The \textbf{q}-summed spectra and their temperature dependences are virtually identical for nearly all of the samples with different dopants, save for the temperature dependence of the Ti-doped sample's spectrum, which we attribute to being affected by a large number of free charge carriers. Comparing with magnetization measurements for different temperatures and dopings likewise did not show a clear correlation between the RIXS spectra and the magnetic ordering states. To clarify the excited states, we performed spin multiplet calculations which were in excellent agreement with the RIXS spectra over a wide energy range and provide detailed electronic descriptions of the excited states.  The implications of these findings to the photoconversion efficiency of hematite photoanodes is discussed.              

\end{abstract}

\maketitle

\section{Introduction}
\label{sect:intro}

Hematite ($\alpha$-$\rm Fe{_2}O_{3}$) and related iron oxides are central to several diverse fields of fundamental science \cite{Morin50,Dzyaloshinsky58,Moriya60,Kupenko19,Catling03,Liang17,Mikhaylovskiy20} and developing potential technologies, including spintronics \cite{Sulymenko17,Lebrun18,Han20} and solar fuels \cite{Sivula11}, to cite a small sample of the numerous significant works.  In the latter application, hematite acts as a photoanode in a photo-electrochemical cell, which absorbs light to generate electron-hole pairs which feed electrochemical ``water-splitting'' reactions to generate hydrogen fuel. Hematite's $R\overline{3}c$ corondum structure \cite{Pauling25,Blake66} can be pictured as hexagonal, with double Fe layer basal planes separated by oxygen atoms, which form tilted octahedra of ligands around each Fe atom. The valence of the Fe atoms in hematite is 3$d^5$, which form a high-spin ground state in accordance to Hund's rule. Their spins are antiferromagnetically ordered along the hexagonal c-axis, but undergo the Morin transition involving a 90$^{\circ}$ rotation between two ordered states at $T=T_M$, with a canting effect producing an additional weak in-plane ferromagnetism for $T>T_M$. \cite{Morin50,Dzyaloshinsky58,Moriya60}. In bulk, undoped hematite, $T_M$=265 K.
Hematite is widely considered to be a ``charge-transfer insulator"~ \cite{ShermanWaite85,Fujimori86,Armelao95,Uozumi97,Fujii99,Pisarev09}, where the lowest energy excitations resulting in mobile charge carriers and visible light photoconductivity are ligand-to-metal charge transfer (LMCT). Estimates of the minimum LMCT peak energy have varied anywhere from 2 eV to over 5 eV. \cite{Armelao95,Pailhe08a,Piccinin20,Marusak80,Debnath82,Zubov81,Pisarev09}. The lowest energy optical absorption features in hematite above 1 eV have been attributed to re-configurations of the $3d$ electrons under the electric field of the surrounding oxygen ligands, or ligand field (LF) excitations \cite{Morin54,Bailey60,TandonGupta70,Marusak80,ChenCahan81,Zubov81,ShermanWaite85}. Coupling between the dipole excitation of LF transitions to inter-site spin interactions \cite{Halley65} in hematite has been demonstrated through various works \cite{Pisarev69,Zubov81,Rossman96,Chen12,Ivantsov20,Mikhaylovskiy20}, and also coupling with phonons \cite{Marusak80}. LF transitions, which are localized and do not generate free charge carriers, can place significant limits on achievable photoconversion efficiency \cite{Grave21,Piekner21}. Moreover, comparisons of photocurrent and optical absorption suggested that LMCT and LF excitations in hematite coexist in the optical absorption spectrum at the same energies \cite{Kennedy78,Hayes16,Grave21,Piekner21,Grave22}, making the task of disentangling and identifying each separately a challenge. Complimentary spectroscopies can therefore play an important role in characterizing the excitations. For example, a recent transient absorption experiment at the Fe $M$-edge, combined with spin-multiplet calculations demonstrated that the main optical absorption feature at 3.1 eV is of predominately LMCT character \cite{Vura-Weis13}. In this paper we study electronic transitions in hematite with resonant inelastic x-ray scattering (RIXS) at the Fe $L$-edge, which is ideally suited for elucidating the LF excitations.\\ 

RIXS is an element-specific, photon-in, photon-out spectroscopy comprised of a core-level absorption, followed by a core-level emission, but with the valence electrons left in an excited state for inelastic processes \cite{Kotani01}. Since the Fe 2$p\rightarrow$~ Fe 3$d$ absorption step of the  $L$-edge RIXS process directly injects an electron into the 3$d$ orbital, and similar (but reversed) for the subsequent emission,  it inherently resonates with LF transitions.  Duda et al, in their early $L$-edge RIXS study of hematite \cite{Duda2000}, observed prominent features at energies corresponding to those in the optical absorption spectrum \cite{Marusak80}, and associated them with LF excitations. Since the core-hole is only an intermediate state, RIXS does not suffer from core-hole related life-time broadening or energy renormalization like x-ray absorption spectroscopy (XAS) does.  More recent $L$-edge RIXS studies of hematite include those by Ye et al. \cite{Ye18}, who observed dependence of relative intensities on sample preparation method, and Miyawaki et al. \cite{Miyawaki17}, who focused on the magnetic circular dichroism of RIXS (RIXS-MCD) in hematite to elucidate the Dzyaloshinkskii-Moriya interaction responsible for the spin canting. RIXS has recently been used to probe hematite's electronic structure response to a light pulse on ultrafast timescales. \cite{Ismail20}, demonstrating one of the first pump-probe transient RIXS studies. An additional capability of RIXS is that, unlike for visible wavelengths, x-ray photons can transfer a significant amount of momentum into electronic excitations, denoted by the vector \textbf{q} referenced to a reciprocal lattice point of an oriented crystal. The evolution of the RIXS energy loss spectra with \textbf{q} could offer clues as to energy vs. momentum dynamics of the states involved in the excitation. A spectral feature whose energy appears to shift with \textbf{q} could indicate a dispersive band or propagating state \cite{Hasan01}, and conversely a stationary peak would imply non-propagating states with high effective mass, as expected for a LF excitation \cite{Sala11} in particular.  While the dispersive case can be subject to misinterpretation (for example, from an intensity-related effect), for an observation of a \textbf{q}-independent excitation, it is likely that the excitation would indeed not involve dispersive bands, thus identifying it with increased certainty as a LF or other flat-band excitation.\\


In this paper we present Fe $L_{3}$-edge RIXS spectra heteroepitaxial hematite thin films, measured for several fixed \textbf{q}, at both room temperature and at $T$ = 14 K.  The samples included undoped and variously doped (both p-type and n-type) to the $\sim$1\% level, which is comfortably below solubility limits but still can boost performance of photoanodes.  A striking similarity in the spectra and their temperature dependence was observed between the majority of differently doped samples we measured, and a strong temperature dependence of the sub-bandgap $\sim$1.4 eV peak was observed, analogous to the feature in the optical spectrum \cite{Marusak80}, and other temperature dependent features. We conclude that most of the observed spectra are composed of LF excitations, based on the mostly flat dispersion and also supported by spin-multiplet calculations. The latter showed excellent agreement with the measured spectra, thus providing detailed state information for the main excitations, which we present as electron density plots. These also resemble the ``non-mobile'' part of the optical absorption spectrum in hematite photoanodes in the visible region, which has been empirically extracted from combined optical absorption and photocurrent spectra \cite{Grave21,Piekner21}, and can have a profound effect on photoconversion efficiency.

\section{Methods}
Heteroepitaxial, single orientation, hematite thin films were deposited on sapphire substrates using pulsed laser deposition, according to the procedure of Grave et al.  \cite{Grave16}. Lattice constants and film thicknesses were extracted by x-ray diffraction analysis, which is specified in the Supplemental Information (SI) \cite{SI} (see also references \cite{Kraus96,Lucht03} therein). Initial RIXS measurements were performed at the qRIXS endstation 8.0.1.3 (Berkely Lab, California) \cite{Chuang17}, Advanced Light Source (ALS), Lawrence Berkeley National Laboratory (LBNL) having an energy loss resolution of $\sim$450 meV, followed by higher resolution (150 meV) measurements at BESSY II synchrotron, U41-PEAXIS beamline (Helmholtz Zentrum Berlin, HZB) \cite{Schulz20}, conducted at both room temperature and $T$ = 14 K. At the ALS qRIXS beamline we measured undoped and 1\% (cation \%) Ti-doped c-axis oriented hematite films, $\sim$150 nm thick hematite samples. We also measured an a-axis sample, which is presented in the SI. For the PEAXIS run, we prepared four samples, each with $\sim$100 nm hematite layer thickness and all of them with c-axis orientation (to within 0.02$^{\circ}$): undoped, 1\% Ti-doped, 1\% Sn-doped, and 1\% Zn-doped. Scattering was in $\theta$-2$\theta$ mode, such that the scattering vector \textbf{Q}=(\textit{h} \textit{k} \textit{l}) was along the surface normal. The scattering plane was horizontal, as was the incident x-ray polarization. For the c-axis samples, the momentum transfers \textbf{q} were along the crystallographic (0 0 1) direction, using hexagonal notation. A small note on notation: the momentum transfer \textbf{q} to the electron is related to the scattering vector \textbf{Q} by \textbf{q}=\textbf{Q}--\textbf{G}. \textbf{G} is an integer Miller index, which is (0 0 1) for the c-axis samples in this work. \textbf{q} may therefore be considered the fractional part of \textbf{Q}, and the two terms may be used interchangeably depending on context. For each energy loss spectrum, the incident energy ($E_i$) was fixed to a value relative to the Fe $L$-edge absorption peak to set resonance condition. The majority of the PEAXIS measurements were for $E_i$ set to the main $L_{3}$ absorption peak energy, as described in the SI in greater detail (see also references \cite{Gota00,Dallera97,ChabotCouture10,Wang20}). The energy loss scales were zeroed according to the elastic line positions, and self-absorption corrections were applied to the intensity, as presented in the data processing section of the SI. Fitting of each spectrum to a sum of Gaussian components is also described in the SI.\\ 
 
To gain insight into the features, a cluster model was applied to the Fe$^{3+}$ RIXS simulations. The many-body Hamiltonian of the cluster model is solved using the Quanty program \cite{Haverkort12}, where the Coulomb multiplet interaction, the spin-orbit coupling on the Fe site, and the charge transfer between Fe 3$d$ and O 2$p$ orbitals are included in the calculation of the 2$p$3$d$ RIXS spectrum \cite{deGroot05}, which is given by the Kramers-Heisenberg formula \cite{Kramers25}.  The experimental scattering geometry was taken into account. The LMCT calculation considers the interaction between $d^n$ and $d^{n+1}\bar{L}$ configurations using the single Anderson impurity model \cite{Anderson61,Green14}, where $\bar{L}$ denotes an empty ligand state. This describes the transfer of an electron from the ligand valence band to cation while still preserving the symmetry and spin \cite{deGroot05}, and can be modeled using the  LMCT energy parameter $\Delta$ and the electron hopping integral $V$. Multi-electron interactions are coded in terms of Slater integrals and spin-orbit coupling energies. The effective spin exchange interaction was set to 0.1 eV, which is approximately consistent with the magnon energy measured by neutron scattering \cite{Samuelsen70}, and Raman scattering \cite{Chen12} 
The model parameters were tuned so as to produce excitation energies and spectra in reasonable accordance with both the observed XAS and RIXS spectral excitations.  Further details of the model calculations are given in the SI (see, also, references \cite{Vercamer16,Cramer91,Haverkort10} therein).\\  
 
To characterize the magnetic state of the variously doped thin films, and in particular to determine $T_M$, magnetization vs. temperature ($M$ vs. $T$) measurements were performed using a Quantum Design MPMS3 system. The actual 1x1 cm$^2$ samples used during the RIXS measurement at BESSY were too large for the SQUID magnetometer, so smaller area 0.5x0.5 cm$^2$ pieces were cut from the same original larger (3x1.5 cm$^2$) films from which the RIXS samples were diced. Probing three points, separated by a cm each, along the original large sample with x-ray reflectometry (details in the SI) showed thickness variations of 2 nm or less, suggesting that the thickness-dependent magnetic properties of the smaller thin film samples should be the same as for the larger RIXS samples. The samples were fixed vertically on a holder with cryogenic varnish (General Electric 7031), with their surface facing horizontally, such that the magnetic field was parallel to the basal plane (ie. perpendicular to the c-axis).  The initial step in the magnetization measurement procedure was to raise the temperature well above 300 K (typically 350 or 400 K), and at that temperature hold a 2T field for two minutes so as to align the weakly ferromagnetic domains.  Then the field was removed and $M$~vs.~$T$ measured down to $T$=10 K.

\section{Results and Discussion}
 
 \begin{figure}[hbp]
 	\centering
 	\vspace*{-3mm}
 	\epsfig{file=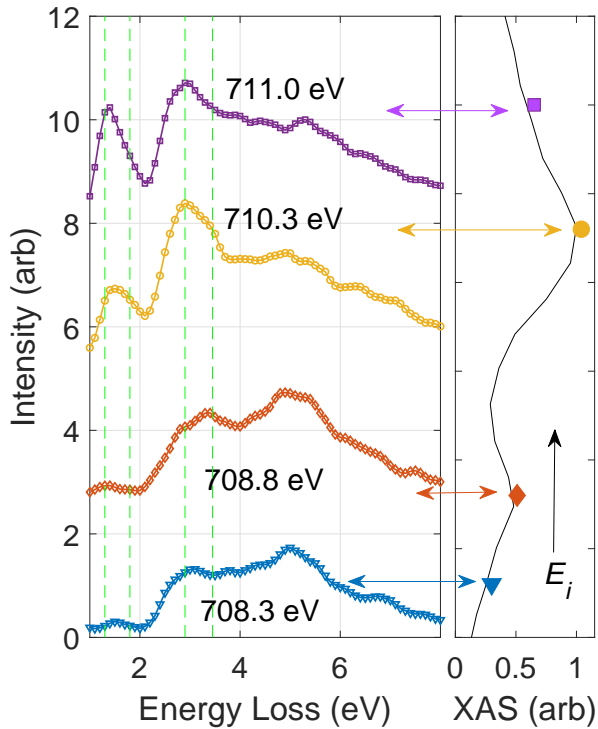,height=4.0in,keepaspectratio}
  	\vspace*{-7mm}
 	\caption{Energy loss scans for different incident energies about the absorption peaks, for the 1\% Ti-doped sample measured at the qRIXS station at ALS, for T=300 K and \textbf{Q}=(0 0 0.9).  The spectra are offset along the y-axis so that their positions roughly correspond to their respective $E_i$'s relative to the XAS spectrum plotted in the right panel. The magnitudes of the spectra scaled for clarity of their shapes.}
 	\label{fig:Fig1}
 \end{figure}


\begin{figure*}[htbp]
	\centering
	\vspace*{-3mm}
	\epsfig{file=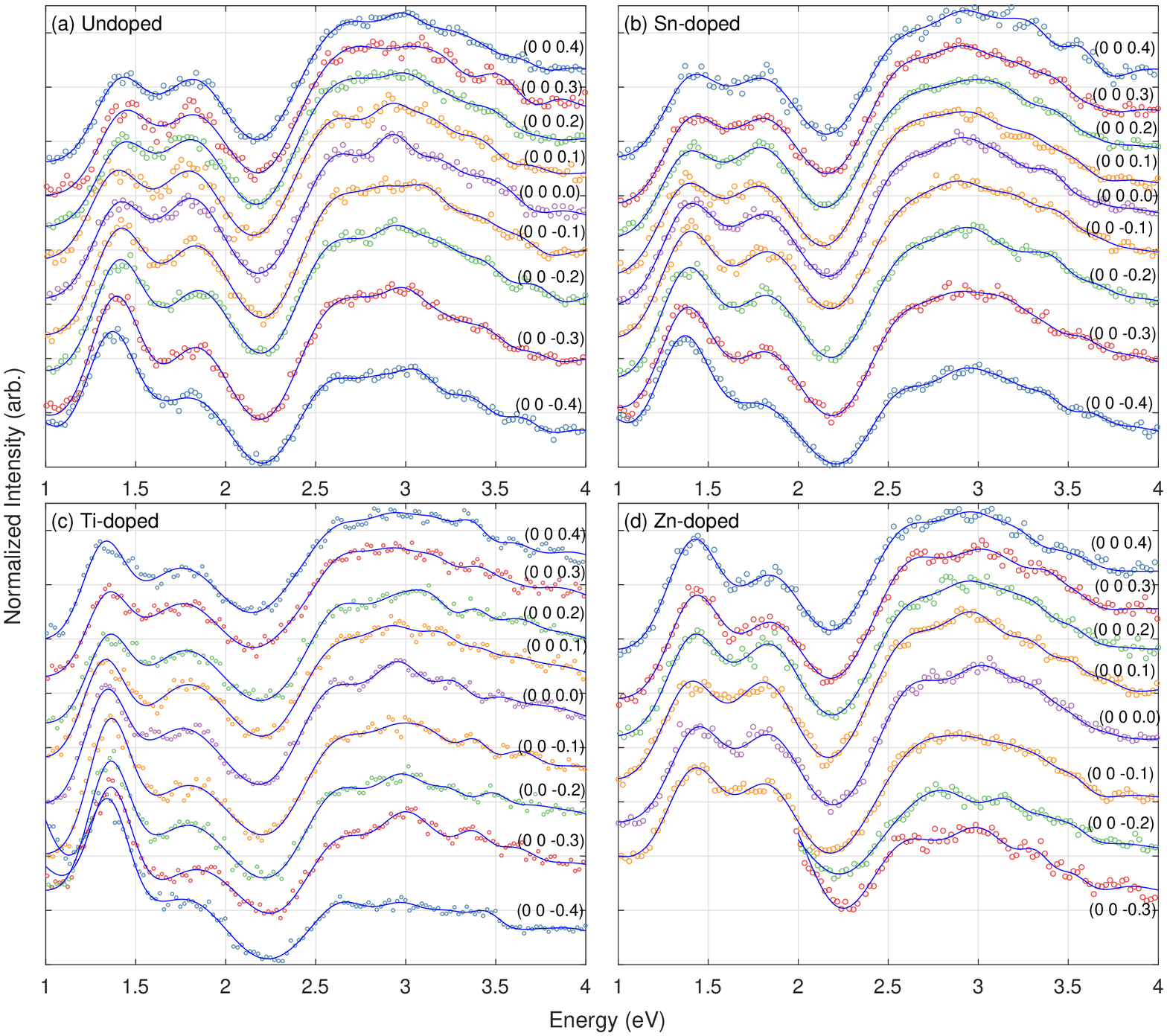,height=6.2in,keepaspectratio}
	\vspace*{-8mm}
	\caption{\textbf{q}-resolved spectra of the $\sim$100 nm thick hematite films measured at the PEAXIS beamline, for T=14 K, for (a) undoped, (b) 1\% Sn-doped, (c) 1\% Ti-doped and (d) 1\% Zn-doped. The Zn-doped spectra at negative q was truncated at low energy because of a high background from the  quasi-elastic tail. The incident energy was $E_i$=710.3 eV.  For clarity, the intensities were normalized so as to be approximately on the same viewing scale, and each curve displaced along the y-axis, therefore the y-axis has arbitrary units, which are not shown. The curves are labeled by \textbf{q}=(0 0 $\delta$) offset from the \textbf{Q}=(0 0 1) position. The solid lines correspond to fitted curves, as described in the main text.}
	\label{fig:Fig2}
\end{figure*}

A survey of the $E_i$-dependence of the energy loss spectra, measured for the Ti-doped sample at the qRIXS beamline is shown in Fig.~\ref{fig:Fig1}(a). The spectra show features at energies common to different $E_i$'s, including those indicated by the dashed vertical lines at $\sim$2.9 eV and $\sim$3.4 eV, whose local peak energies are identical, but whose relative magnitudes of resonance change according to $E_i$.  Excitations are also observed in the 1.3--1.8 eV range, below the $\sim$2 eV optical bandgap of hematite. This corresponds to the 1.4 eV and 1.8 eV peaks observed as relatively weak features in optical absorption, and will be more clearly resolved in the higher resolution measurements below. The spectral shapes in Figure 1, and their $E_i$ dependence, closely resemble those measured by Miyawaki et al. in their recent magnetic circular dichroism (MCD) RIXS study \cite{Miyawaki17}.\\

Panels (a)-(d) of Figure 2 show the \textbf{q}-dependence of the differently doped samples measured at the PEAXIS beamline, for $E_i$=710.3 eV fixed at the Fe $L_{3}$ absorption peak and $T$ = 14 K. The 150 meV energy resolution resolves (at least) the two sub-bandgap peaks at $\sim$1.4 and $\sim$1.8 eV.  Fits to multiple Gaussians, shown as solid lines in Figure 2 and described in the SI, suggest that the main envelope of the spectrum does not appreciably change with \textbf{q}. Although this does not rule out weak energy-dispersive features, it does imply that most of the spectral contributions are non-dispersive, and are likely LF excitations. Individual spectra for room temperature, as well as 2D plots showing possible energy dispersions (including measurements from both PEAXIS and qRIXS beamlines, and comparisons between them) including a possibly dispersive branch above 3 eV, and a slight curvature observed in the intensity edge around 2.5 eV, are presented and discussed in more detail in the SI (see, also, references \cite{Schlappa12,Kim12,Bisgoni15,Fumagalli20,Goodenough71,Wohlfeld13} therein).  While these possible dispersions might be potentially indicate propagating modes (not necessarily charge-carrying), the associated intensities are too weak to be conclusive, and we focus instead on the LF excitations which are intense at the Fe $L$-edge.  Complimentary RIXS measurements at the hard x-ray Fe $K$-edge, where charge transfer excitations are expected to be relatively stronger than LF intensities \cite{Ament11}, could be a promising route to investigate dispersive charge transfer features in hematite.\\

   \begin{figure*}[hbp]
  	\centering
  	\vspace*{-3mm}
   \epsfig{file=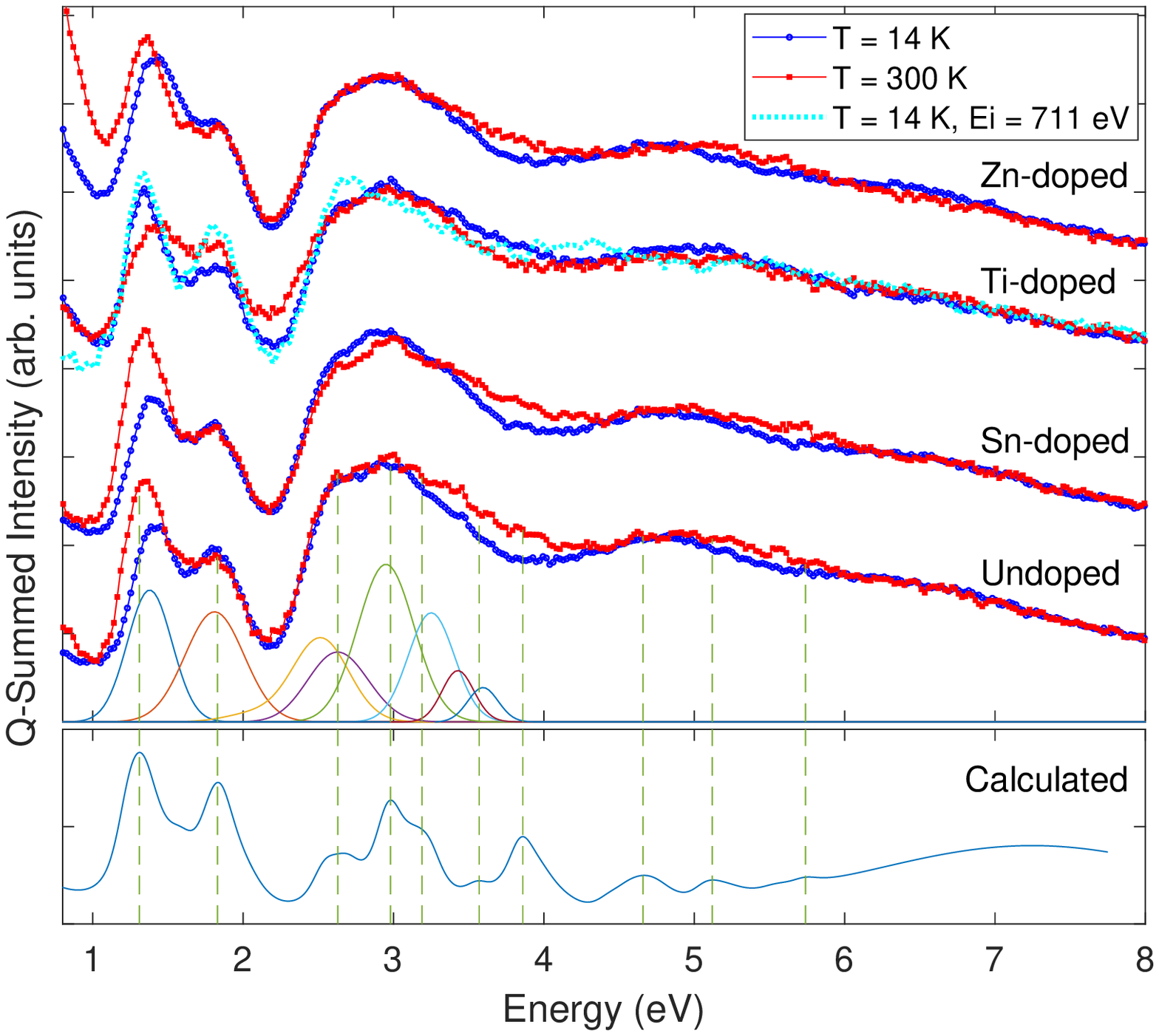,height=6.1in,keepaspectratio}
  	\vspace*{-8mm}
  	\caption{Top panel : \textbf{Q}-integrated spectra by summing the \textbf{Q}-resolved spectra in the range \textbf{Q}=(0 0 0.6)-(0 0 1.4), for $E_i$=710.3 eV. measured at T=14 K (blue) and T=300 K (red). The cyan-coloured curve is for Ti-doped, $E_i$=711.0 eV, T=14K, summed over the range \textbf{Q}=(0 0 0.8)-(0 0 1.2), discussed later in the text.  The data were normalized so as to overlap in the high-energy region.  The spectra of each sample are displaced along the y-axis.  The Gaussian lineshapes plotted below the measured spectra represent typical fitted components up to 4 eV energy loss, obtained from averaging over \textbf{Q} the fitted parameters of the T=14K spectra of the Sn-doped sample. The vertical green lines correspond to the energies of the simulated peak intensities shown on the bottom panel. Bottom panel : Simulated intensities. The vertical dashed green lines were placed at the approximate local maxima.}
  	\label{fig:Fig4}
  \end{figure*}

An overall picture of the temperature- and sample-dependence of the spectra may be viewed in a compact form by summing the individual \textbf{q}-resolved spectra for each sample/temperature. The resultant ``q-integrated" spectra are plotted for each sample in the top panel of Figure 3, with T=300 K in red, and T=14 K in blue. All of the spectra were normalized so as to have the same intensity in the high-energy region above 7 eV, where the intensity appears to fall off linearly. Thus normalized, the majority of the spectra at lower energy in Figure 3 line up almost exactly, both between temperatures and samples (with the exception of the Ti-doped sample, which will be discussed in more detail separately).  It is therefore reasonable that the portions of the spectra measured at $T$ = 14 K and $T$ = 300 K respectively, that do \textit{not} overlap should be considered to be the truly temperature-dependent features. Of these, three stand out as obvious and common to the majority of samples: the $\sim$1.4 eV peak, which increases intensity and red-shifts at higher temperature; the broad 3-4 eV region which increases in intensity with higher temperature; and a broad peak at $\sim$4.7 eV, whose peak energy blue-shifts to $\sim$5.2 eV at higher temperatures. Simulated peak intensities after optimizing the parameters in the spin multiplet model are shown in the lower panel of Figure 3. Details regarding the parameters and a Tanabe-Sugano diagram can be found in the SI. In addition to lining up with the 1.4 and 1.8 eV peaks, the calculated peak energies are in remarkable agreement with other key aspects of the spectra, including the large gap between $\sim$1.8 and $\sim$2.5 eV, and at least qualitative agreement of the simulated intensity features with the q-averaged Gaussian fit components in the upper panel of Figure 3, as shown by the vertical dashed lines connecting the upper and lower panels. Higher energy features are also to some degree captured by the calculations. \\

\begin{figure}[!htb]
	\centering
	\vspace*{-3mm}
	\epsfig{file=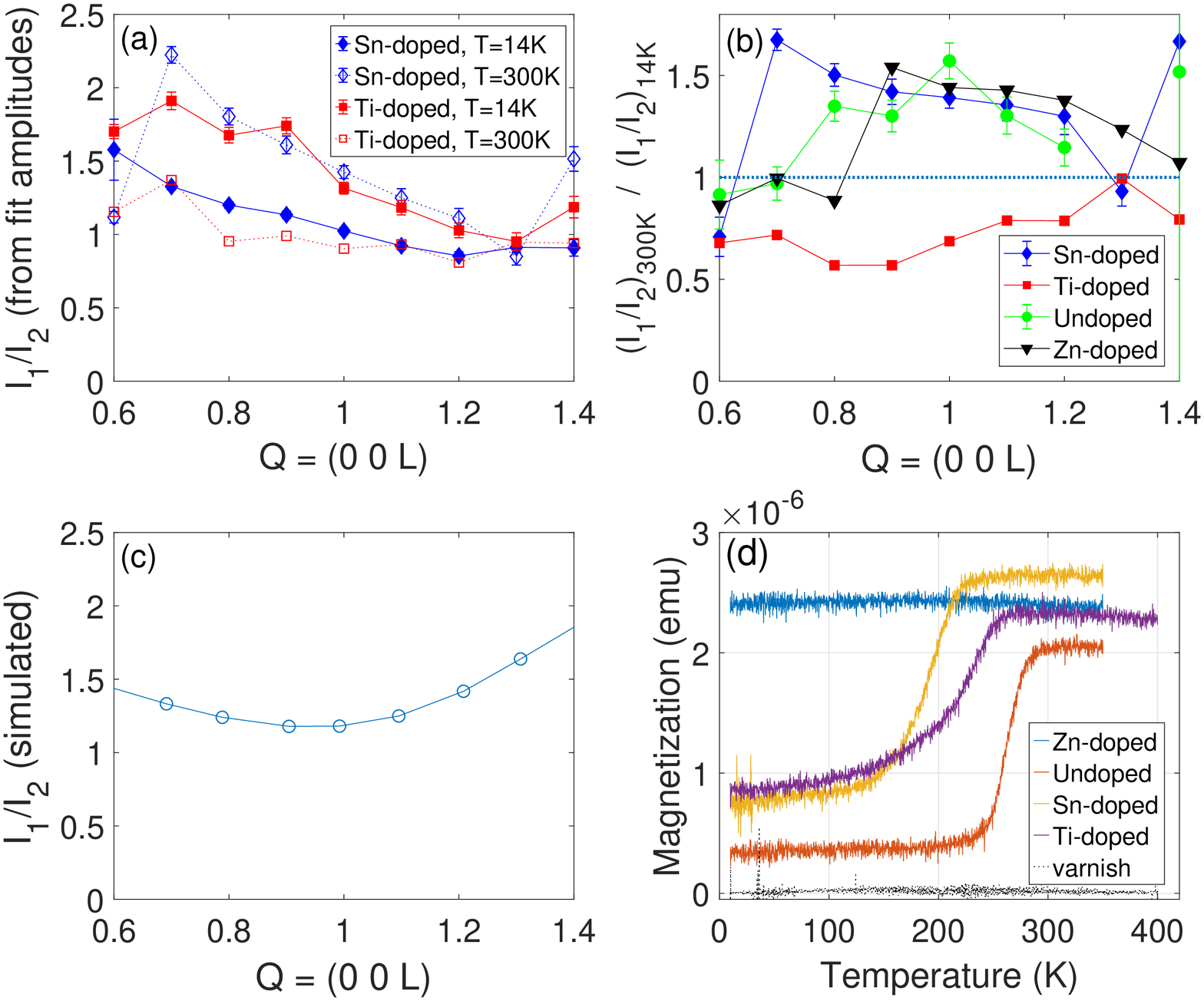,height=3.0in,keepaspectratio}
	\vspace*{-8mm}
	\caption{(a) Ratio of fitted Gaussian peak heights $I_1$ ($\sim$1.4 eV) to $I_2$ ($\sim$1.8 eV) as a function of momentum transfer, for the Sn-doped (blue diamonds) and Ti-doped (red squares)  samples at T=14 K (filled symbols) and T=300 K (empty symbols). (b) Ratio of the $(I_1/I_2)$ ratio at T=300 K to that at T=14 K, versus momentum transfer \textbf{Q}, for the Sn-doped, Ti-doped, undoped  and Zn-doped samples. (c) $I_1$:$I_2$ ratio obtained from the simulations  (d) $M$ vs. $T$ under conditions of zero-field cooling, for the differently doped equivalent films, as described in the text.}
	\label{fig:Fig5}
\end{figure}

 \begin{figure*}[htbp]
	\centering
	\vspace*{-3mm}
	\epsfig{file=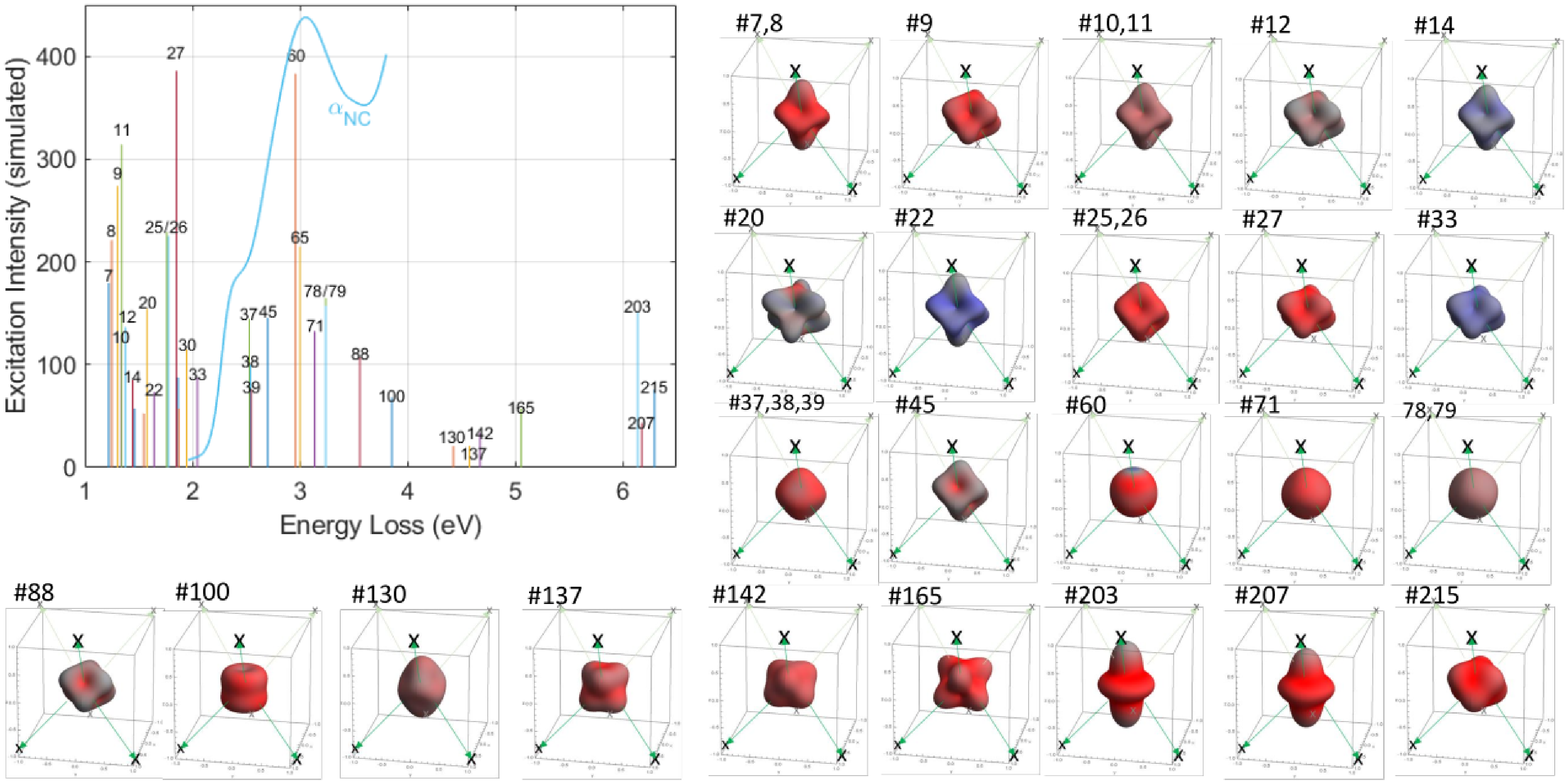,height=4.15in,keepaspectratio}
	\vspace*{-8mm}
	\caption{The highest intensity excitations calculated for \textbf{Q}=(0 0 1.1), plotted as a stick figure, with each excitation labeled according to state number, as described in the text. The total density of the combined Fe 3\textit{d} electrons of the simulated excited states is plotted according to each state number. In these electron density plots, the crystallographic c-axis is vertical.  The distance scale is arbitrary, but the approximate oxygen angular locations are marked with x's, whose sizes indicate relative position normal to the page (ie. larger are more in front, smaller and dimmer behind).  They form a triangle on the bottom surface of the cuboid and an inverted triangle on the top surface. The view of the far-bottom oxygen position may be fully or partially obstructed by the Fe 3d density image. The color shading of the electron density plots reflects the expectation of the projection of the combined spin along the c-axis relative to the ground state alignment determined by magnetic ordering; bright red is for the ground-state direction, while fully blue corresponds to all spins flipped.  Close-lying states with similar energies and density plots were labelled jointly. Owing to space constraints, the density plot of state 30 was omitted, but its orbital profile is similar to state 33, only with lower expectation for $S_z$. Likewise, state 65 is similarly isotropic as state 60. The continuous blue curve is the ``non-contributing'' component of optical absorption, obtained from reference [35] as described in the text, scaled to fit the present y-axis.}
	\label{fig:Fig6}
\end{figure*}

The Ti-doped sample appears to have a temperature dependence contrary to that of the other samples, as shown by the energy shift and relative intensity of the $\sim$1.4 eV peak, and the intensity of the 3-4 eV region, in Figure 3.  To characterize this apparent reversal quantitatively, we examine the \textbf{Q}-dependence of the $\sim$1.4 peak intensity $I_1$, normalized by the $\sim$1.8 eV peak intensity $I_2$ which is relatively constant with temperature for the other samples. The intensities were obtained from the fitted Gaussian component amplitudes. Figure 4(a) plots the $I_1/I_2$ trend for the two temperatures, for the Ti-doped (red squares), and Sn-doped (blue diamonds) samples. The shape of the overall trends are similar for both samples and temperatures, and is also qualitatively comparable to the simulations (Figure 4(c)). From afar, the curves for the Sn- and Ti-doped samples in Figure 4(a) look almost overlapping, with a gap separating two pairs of approximately overlapping curves. However, it is the $T$=14 K curve of the Ti-doped that overlaps with the $T$=300 K curve of the Sn-doped, and vice-versa, as if the effect of temperature has been reversed for the Ti-doped sample. To plot all of the dopants together (Figure 5(b)), we divide their respective $T$=300 K curves by their $T$= 14 K curves, which is effectively to plot the increase of the (normalized) 1.4 eV peak for $T$=300 K as compared to $T$=14 K. Near the zone boundary the results vary, but within the central \textbf{Q}$\sim$(0 0 0.9)-(0 0 1.2) range the Zn-doped (black), Sn-doped (blue), and undoped (green), approximately overlap, showing an increase of the order of $\sim$40\% at $T$=300 K. Plotted this way, the Ti-doped (red) curve appears to be roughly a mirror image of the other curves about the unity line. Reviewing possible causes for this apparently opposite temperature dependence, it should be mentioned that there was a beam interruption during the low-temperature measurement of the Ti-doped sample. One concern in cases of beam interruption would be a shift in incident energy $E_i$ because of monochromator temperature. However, to our knowledge the system had sufficiently recovered by the time the measurements were resumed. A detailed account of the experimental conditions (elastic line position on the CCD, calibration tables, etc.) and data relating to the measurement of this sample is given in the SI, and shows no indication of an $E_i$-related cause. In Figure 3 we include a spectrum measured for a different $E_i$, where the 1.4 eV peak is enhanced (and thus may produce a similar effect if $E_i$ was thus shifted): the $T$=14 K spectrum for $E_i$=711 eV (cyan line). While the 1.4 eV intensity is indeed enhanced, other parts of the spectra also significantly change at this $E_i$. But since the basic shape of the $E_i$=710.3 eV Ti-doped spectra is the same as the others, and the spectra for both temperatures almost exactly overlap in the 2.5-3 eV range, we can rule out a shift in $E_i$. Another possible experimental check could be to measure azimuth angle dependence at different temperatures, but this has not been explored, and the anomalous temperature behavior of the Ti-doped sample remains an open question. We do note, however, that in a previous comparison of  hematite photoanode samples doped at the 1\% level with different dopants \cite{Malviya16}, the Ti-doped sample stood out as having a much higher carrier concentration as determined from  Mott-Schokkty analysis, an order of magnitude higher than for the others. This, together with the present result, could suggest a coupling effect of free electrons to the temperature-dependent features in the RIXS spectrum.  \\

$M$~vs.~$T$ measurements, plotted in Figure 4(d), were used to check for any correlations between the RIXS spectra and the sample's magnetic ordering state. In thin hematite films, $T_M$ is known to be sensitive to film thickness \cite{Park13,Shimomura15,Mibu17} and dopant \cite{Ellis17}, so a comparison of the magnetic state of variously doped thin films against their respective RIXS spectra (measured at $T$=14 K and $T$=300 K) should indicate whether or not the RIXS spectrum appears to be affected by whether $T>T_M$ or not.  The magnetic state as probed by soft x-rays, having a penetration depth of approximately 20-30 nm, has been shown to be consistent with the magnetic state throughout the bulk of the film. $T_M$ measured by x-ray dichroism at the Fe $L_{3}$ edge for an undoped 150 nm thin film \cite{Ellis17}, for example, was the exact interpolation of the $T_M$ vs. thickness curve published by Shimomura et al. \cite{Shimomura15}, which was based on magnetization measurements. All of the samples save Zn-doped underwent a Morin transition, indicated by a step in the magnetization in Figure 4(d). This is consistent with the previous x-ray dichroism study \cite{Ellis17}, in that a similar 1\% Zn-doped sample (150 nm hematite film) was also found to always be in the Morin state down to low temperatures. Both the Ti-doped and Sn-doped samples are in the Morin state at $T$=300 K, and well into the low-$T$ antiferromagnetic state at $T$=14 K, so this cannot explain the differences in ($I_1/I_2$) for the two samples in Figure 4(a). Furthermore, the Zn-doped sample remains in the Morin state even at low temperature, yet it nevertheless has mostly similar temperature dependence of the $I_1$/$I_2$ ratio as the Sn-doped and undoped samples, as well as exhibiting the same energy shift in the $\sim$1.4 eV peak. There may be subtle effects upon close inspection of the magnetization and/or RIXS spectra, such as the shallower slope of the Morin transition for the Ti-doped sample in Figure 4(d), or the zone-boundary spectra of the Zn-doped sample at low temperature (Figure 2(d)), which looks similar to that of the Ti-doped (Figure 2(c)).  However, on first view, the doping/temperature dependence of the RIXS spectra and the magnetic state do not clearly correlate. \\

While the RIXS data and simulated spectra in Figure 3 show only a relatively few main features, the five electrons in the Fe 3\textit{d} orbitals give rise to a multitude of possible multiplet states, numbering in the hundreds.  A list of states and their expectation values for angular momentum and related observables is made available in the SI.  To visualize the main features, we select only the states having the highest calculated intensities for \textbf{Q}$\simeq$(0 0 1.1) ($2\theta=90^\circ$), as ``sticks'', or zero-broadened peaks showing only energies and relative intensities. These are plotted in Figure 5, and labeled according to their respective state numbers as listed in the SI table. The electron density plots of the combined Fe 3\textit{d} electrons' wave function associated with these states are also plotted in Figure 5, relative to the (approximate) oxygen angular positions marked by ``x''. Because of the empirical parameters used for the spin-multiplet calculations, the length scale in the electron density plots is arbitrary. The color of each density plot is representative of the combined spin component along the c-axis, the $S_z$ expectation value, whereby bright red represents the ground-state spin (i.e. all spins aligned either up or down, depending on the antiferromagnetic sub-lattice of the Fe site), and as the color changes to dimmer red towards blue, the mean spin increasingly changes direction, with bright blue representing all electron spins flipped in opposition to the antiferromagnetic order.\\

The first two rows of density plots in Figure 5, corresponding to the sub-bandgap 1.2--2.1 eV energy range, have a net $t_{2g}$-like character, with indentations or nodes facing the oxygen atoms. To a first order approximation, these may be likened to either $t_{2g}^4e_g^1$ or $t_{2g}^5e_g^0$ type states or mixtures thereof, in qualitative agreement with the calculations of Miyawaki et al. for this energy range. \cite{Miyawaki17}  With increasing energy, the states exhibit a compression along the c-axis while elongating within the basal plane, and/or they increasingly flip spins against the magnetic order. We stress that the density plots do not contain all of the state information; for this we refer to the states list available in the SI. States 9 (1.3 eV) and 27 (1.85 eV), for example, have similar looking density plots, but the latter has somewhat higher expectations for orbital and total angular momentum magnitudes.
The third row of density plots in Figure 5 corresponds to the 2.5--3.3 eV range. Most of these states are more isotropic (similar to spin-flip states below 1 eV, not shown) and have a less obvious $t_{2g}$ character as compared to the first two rows, but have higher expectation values for orbital and total angular momentum magnitudes. State 45 (2.7 eV) retains a net $t_{2g}$-like character, but its low spin magnitude expectation value distinguishes it from the lower-energy states with similar looking density plots. Starting with state 100 (at 3.85 eV) in the bottom row, the electron density profiles take on a more $e_g$-like character, with lobes facing towards the oxygen atoms. The continuous blue curve in Figure 5 plotted in the 2-4 eV region is the ``non-contributing'' (or immobile) part of the optical absorption coefficient which was extracted from a combination of spectroscopic ellipsometry and photocurrent measurements of a hematite photoanode \cite{Piekner21}. The local maxima of the extracted localized spectrum at $\sim$2.5 eV and $\sim$3 eV are consistent with the stick plots of the calculated excitations. Interestingly, the extracted immobile spectrum continues to have appreciable intensity beyond 4 eV, which suggests that higher energy LF excitations which have low intensities in RIXS have higher intensities in optical absorption (which is the reverse case for the sub-bandgap LF excitations). These might correspond to the $e_g$-like states clustered around 4.5 eV in Figure 5, or perhaps excitations with even smaller RIXS intensity. We note that Hayes et al., in their optical absorption study, likewise assigned an LF contribution centered around this energy. \cite{Hayes16}.\\

The last three states (203, 207, 215) shown in Figure 5, above 6 eV, mark the onset of charge transfer states in our calculations. We fixed the charge transfer parameter $\Delta$ to 2 eV, based on the value that Fujimora et al. extracted from their photoemission measurements \cite{Fujimori86}.  However, the authors of that work remarked that the resultant predicted charge transfer excitation energy of above 6 eV (consistent with the present calculation) would be significantly higher than the observed bandgap of hematite, and went on to propose an alternative scenario for the conduction band, such as an Fe 4\textit{s} band. As discussed in relation to the q-dependence plots of Figure 2, it is difficult to experimentally pinpoint the charge transfer excitations from $L$-edge RIXS dispersion plots alone, and RIXS at the metal (Fe) $K$-edge would be a more promising absorption edge for enhancement of LMCT transitions \cite{Ament11}. Nevertheless, close inspection of the T = 14 K spectra of the Zn-doped, Sn-doped, and undoped samples in Figure 3 shows a relatively weak and broad peak centered at $\sim$6.5 eV. Simple fitting of this feature for each Q is presented in the SI, and resulted in a dispersion (albeit with high error-bar) of the order of 100 meV, with an especially smooth dispersion for the Zn-doped sample in the Q=(0 0 1)-(0 0 1.3) range. Therefore this feature can be tentatively assigned to LMCT excitation(s), although it does not preclude the possible existence of LMCT at lower energies.

\section{Summary}

In summary, we have measured the Fe $L$-edge, momentum resolved, RIXS spectra of hematite heteroepitaxial films for different dopants and temperatures. Although possibly dispersive features could be observed in the spectra which could indicate propagating modes of some form, the majority of the RIXS spectra at the Fe $L$-edge peak is composed of excitations to localized, non-propagating LF-type states. This conclusion was confirmed with spin-multiplet calculations, which showed quantitative agreement with the measured energies and intensities, and elucidated the spin and orbital natures of each peak. While much of the normalized RIXS spectral shape exhibited little temperature dependence, certain regions, namely the sub-bandgap peak at 1.4 eV, the 3-4 eV region, and the excitation peaks about 5 eV, showed pronounced temperature dependence. Doping up to the 1\% level with various dopants had little effect on the observed RIXS spectra, with the exception of the Ti-doped sample, whose temperature trend appeared to be an outlier (possibly a result of exceptionally large amount of electrons donated by this dopant), although its general spectral shape was nevertheless very similar to the others. By comparing the temperature/doping dependence of the magnetization with the RIXS spectra, the q-averaged spectra do not appear to be sensitive to the magnetic ordering state (ie. whether low-$T$ AFM or the Morin state with weak ferromagnetism). Rather, a phonon-coupling or temperature-induced lattice or symmetry perturbation effect is likely responsible for the observed spectral weight transfers between excitations, as the temperature is changed, as proposed by Marusak et al. \cite{Marusak80} for the optical case. Although a small amount of doping can have a profound effect on the magnetism in thin film hematite, and likewise photoanode performance by affecting charge transport in the bulk and/or between surface and electrolyte, as well as surface recombination, we show that it has relatively little effect on the LF spectral shape, at least as seen by RIXS.  But, this is not to exclude the possibility that it could nevertheless impact the LF optical transition cross-sections, which could be a subject of future experimental and theoretical investigation.\\

Fe $L$-edge RIXS combined with spin-multiplet calculations is a powerful tool to study ligand field excitations in detail. Further measurements at the $L$-edge could include varying polarization and in-plane \textbf{q} to observe intensity modulations such as in Sala et al \cite{Sala11}, to test the orbital symmetry calculations presented in Figure 5; while the 1.4 and 1.8 eV peaks should exhibit such modulations, it should be less so for the 3 eV peak, which has a more isotropic density profile.  For non-ligand field, more delocalized excitations, a momentum-resolved Fe $K$-edge RIXS is suggested as a complimentary route to study (possibly more mobile) excitations in hematite. Finally, the peak assignments in RIXS could provide an important perspective when interpreting features in the optical absorption spectra for solar energy applications. A recent empirical study (by some of us \cite{Piekner21}) of optical absorption in conjunction with spectrally-resolved photocurrent in hematite, found that the optical absorption resulted in a significant proportion of non-mobile excitations peaked around $\sim$2.5 eV, 3.0 eV, and around 3.6 eV, suggesting a link to the excitations determined in the present RIXS-spin multiplet study shown in figure 5. We hope that such combined information can be helpful for constructing a model that can properly account for the coupling mechanisms leading to the observed cross-sections of optical transitions in hematite and similar materials, which in turn, can lead to more informed designs of photoanodes or for other energy applications.       

\section{Acknowledgments}  

We are grateful to Dr. Anna Eyal, of the Quantum Matter Research Center of the Russel Berrie Nanotechnology Institute (RBNI) Technion-Israel Institute of Technology, for performing the magnetization measurements. This research used resources of the Advanced Light Source, a U.S. DOE Office of Science User Facility under contract no. DE-AC02-05CH11231. Measurements were also carried out at the U41-PEAXIS beamline at the BESSY II electron storage ring operated by the Helmholtz-Zentrum Berlin für Materialien und Energie. We appreciate attentive technical support of Dr. Maciej Bartkowiak throughout the PEAXIS beamtime, and of Dr. Xuefei Feng (presently at USTC) during the experiment at the qRIXS beamline. We also thank Dr. Roel van de Krol for encouragement. The research was supported by the Israel Science Foundation through the PAT Center of Research Excellence and grant no. 1867/17. The research leading to this result has also been supported by the project CALIPSOplus under the Grant Agreement 730872 from the EU Framework Programme for Research and Innovation HORIZON 2020. Part of this research was carried out within the Helmholtz International Research School ``Hybrid Integrated Systems for Conversion of Solar Energy'' (HI-SCORE), an initiative co-funded by the Initiative and Networking Fund of the Helmholtz Association. R.-P. W acknowledges funding by the German Ministry of Education and Research (BMBF) grant no. 05K19GU2. A.R. acknowledges the support of the L. Shirley Tark Chair in Science.

\end{document}